\documentclass[prl,twocolumn,showpacs,superscriptaddress,floatfix]{revtex4}

\usepackage{graphicx}
\usepackage{amsmath}
\usepackage{amssymb}

\def\myref#1{(\ref{#1})}
\def\be{\begin{equation}}
\def\ee{\end{equation}}

\begin{document}

\title{Coherent ``metallic'' resistance and medium localisation
in a disordered 1D insulator}

\author{Martin \surname{Mo\v{s}ko}}
\email{martin.mosko@savba.sk}
\affiliation{Institute of Electrical Engineering, Slovak Academy of
Sciences, D\'{u}bravsk\'{a} cesta 9, 841 04 Bratislava, Slovakia}
\affiliation{Institut f\"{u}r Schichten und Grenzfl\"{a}chen,
Forschungszentrum J\"{u}lich GmbH, 52425 J\"{u}lich, Germany}

\author{Pavel \surname{Vagner}}
\affiliation{Institut f\"{u}r Schichten und Grenzfl\"{a}chen,
Forschungszentrum J\"{u}lich GmbH, 52425 J\"{u}lich, Germany}
\affiliation{Institute of Electrical Engineering, Slovak Academy of
Sciences, D\'{u}bravsk\'{a} cesta 9, 841 04 Bratislava, Slovakia}

\author{Michal \surname{Bajdich}}
\affiliation{Institute of Electrical Engineering, Slovak Academy of
Sciences, D\'{u}bravsk\'{a} cesta 9, 841 04 Bratislava, Slovakia}

\author{Thomas \surname{Sch\"{a}pers}}
\affiliation{Institut f\"{u}r Schichten und Grenzfl\"{a}chen,
Forschungszentrum J\"{u}lich GmbH, 52425 J\"{u}lich, Germany}

\date{\today}

\begin{abstract}
It is believed, that a disordered one-dimensional (1D) wire with
coherent electronic conduction is an insulator with the mean resistance
$\left< \rho \right> \simeq e^{2L/\xi}$ and resistance dispersion
$\Delta_{\rho} \simeq e^{L/\xi}$, where $L$ is the wire length and
$\xi$ is the electron localisation length. Here we show that this 1D
insulator undergoes at full coherence the crossover to a 1D ``metal'',
caused by thermal smearing and resonant tunnelling. As a result,
$\Delta_{\rho}$ is smaller than unity and tends to be
$L/\xi$-independent, while $\left< \rho \right>$ grows with $L/\xi$
first nearly linearly and then polynomially, manifesting the so-called
medium localisation.
\end{abstract}

\pacs{73.23.-b, 73.61.Ey}

\maketitle

Any coherent electron wave in a one-dimensional (1D) wire with
uncorrelated disorder is exponentially localized \cite{Mott-61}. The
resistance $\rho$ of such a disordered 1D wire should therefore
increase exponentially with the wire length, $L$. In an ensemble of
macroscopically identical wires the resistance fluctuates from wire to
wire owing to disorder randomness, so it is natural to study the mean
resistance $\left< \rho \right>$, where the angular brackets denote the
average over the ensemble. Landauer found at zero
temperature~\cite{Landauer-70}
\be\label{tra}
\left< \rho \right>=0.5 \ [\exp(2L/\xi)-1] ,
\ee
where $\xi$ is the electron localisation length (throughout this paper,
$\rho$ is a dimensionless resistance in units $h/2e^2$). He also found
that $\left< 1/\rho \right>$ diverges and noted that $\left< \rho
\right>$ is not representative of the ensemble. Anderson et al.
\cite{Anderson-80} therefore defined the typical resistance $\rho_t
\equiv \exp{\left< f \right>} -1$, where $f=\ln(1+\rho)$ and $\left< f
\right>$ is the ensemble average of $f$. They showed for zero
temperature that $\left< f \right> = L/\xi$, i.e.,
\be\label{rt}
\rho_t=\exp(L/\xi)-1 .
\ee
The variable $\rho$ exhibits giant fluctuations, the dispersion
$\Delta_{\rho} \equiv (\left< \rho^2 \right> - \left< \rho
\right>^2)^{1/2}/ \left< \rho \right>$ can be shown to grow as
\cite{Mello-87}
\be\label{disprho}
\Delta_{\rho}  \simeq \exp(L/\xi).
\ee
However, $\Delta_{f} \propto 1/\sqrt {L}$, which means that the
variable $f$ self-averages and $\rho_t$ is representative of the
ensemble~\cite{Anderson-80}.

The above formulae hold also for a quasi-1D wire with many 1D
channels~\cite{Thouless-77,Anderson-81}. Thus, for $T=0$~K and $L>\xi$
any 1D wire is insulating, i.e., both the resistance and fluctuations
grow exponentially with $L/\xi$.

Thouless \cite{Thouless-77} argued that at nonzero temperatures the
exponential rise with $L$ would not be apparent since the inelastic
collisions would cause electrons to hop from one localised state to
another one. This would cause $\rho \propto L$ and the 1D localisation
would be manifested solely by a characteristic temperature dependence
of hopping. All this would happen if the electron transport time
through the wire exceeds the inelastic scattering time.

At even higher temperatures, the electron coherence length $L_{\phi}$
would become smaller than $\xi$. This would give rise to the
``metallic'' resistance ($\rho \simeq L/\xi$) and weak localisation
(see e.g. Ref.~\onlinecite{Datta-95}). The crossover between the
``metallic'' resistance and hopping, first predicted in
Ref.~\onlinecite{Thouless-77}, has been observed in the quasi-1D wires
\cite{Gershenson-97}.

Therefore, the disordered 1D insulator described by eqs. (\ref{tra} -
\ref{disprho}) is believed to exist at appropriately low temperatures,
i.e., without hopping and at full coherence \cite{Datta-95,Imry-02},
and the crossover to the 1D ``metal'' seems to occur only if inelastic
collisions are present.

We show that the crossover to another 1D ``metal'' occurs at full
coherence due to the thermal smearing and resonant tunneling. It
results in the wire resistance rising with $L/\xi$ first nearly
linearly and then polynomially owing to the medium localisation. In
addition, $\Delta_{\rho}$ becomes smaller than unity and tends to be
$L/\xi$-independent.

The insulator - metal transition at $T=0$~K has recently been found in
a 1D solid with specially correlated disorder~\cite{Carpena-02}. In our
work the disorder is uncorrelated and the reported insulator -
``metal'' crossover, albeit also coherent, is driven by low nonzero
temperatures.

We consider a 1D wire with disorder described by potential $V(x)=\sum
_{j=1}^N \gamma\delta(x-x_j)$, where $\gamma\delta(x-x_j)$ is the
$\delta$-shaped impurity potential of strength $\gamma$, $x_j$ is the
$j-$th impurity position randomly chosen along the wire, and $N$ is the
number of impurities. If the 1D electrons tunnel through disorder
coherently, the electron wave function $\Psi_k(x)$ is the solution of
the tunneling problem
\begin{equation} \label{e3}
\left[-\frac{\hbar^2}{2m}\
\frac{ d^2}{dx^2}+V(x)\right] \Psi_k(x)=\varepsilon \, \Psi_k(x),
\end{equation}
\begin{equation} \label{e4a}
\Psi_k\left(x\rightarrow0\right)=e^{ikx}+r_k e^{-ikx}, \ \
\Psi_k\left(x\rightarrow L\right)=t_k e^{ikx},
\end{equation}
where $\varepsilon = \hbar^2k^2/2m$ is the electron energy, $k$ is the
wave vector, $m$ is the effective mass, $r_k$ is the electron
reflection amplitude, and $t_k$ is the electron transmission amplitude.
In the boundary conditions~(\ref{e4a}) the electron impinging disorder
from the left is considered. The beginning and end of the wire are set
at $x=0$ and $x=L$, respectively.

We find the amplitude $t_k$ by means of the transfer matrix (TM)
method. The TM of disorder $V(x)$ reads~\cite{Davies-98}

~\vspace{-0.8cm}~
\begin{equation} \label{TotTM}
\mathbb{T} =
\begin{pmatrix}
1/t_k^* & -r_k^*/t_k* \\
-r_k/t_k & 1/t_k
\end{pmatrix}
= \mathbb{T}(x_N) \, \dots\, \mathbb{T}(x_2)\, \mathbb{T}(x_1),
\end{equation}
where
\vspace{-0.5cm}
\begin{equation} \label{TMdeltaX}
\mathbb{T} (x_j) =
\begin{pmatrix}
1-i\Omega/k \, & -i\frac{\Omega}{k} \, e^{-2ikx_j} \\
i\frac{\Omega}{k} \, e^{2ikx_j} & 1+i\Omega/k
\end{pmatrix}
\end{equation}
is the TM of the $\delta$-barrier at position $x_j$ and $\Omega=
m\gamma/\hbar^2$. We evaluate the right hand side of eq.~\myref{TotTM}
numerically.

We then obtain the transmission probability $\mathcal{T} = {|t_k|}^2$
and evaluate numerically the two-terminal conductance
\begin{equation} \label{gT}
G = \int _0 ^{\infty} d\varepsilon \ \left[-
 \frac{ d\text{f}(\varepsilon)}{d\varepsilon} \right]
  \, \mathcal{T}(\varepsilon) ,
\end{equation}
where $\text{f}(\varepsilon)$ is the Fermi distribution. To calculate
the wire resistance, instead of $\rho = 1/G$ we use the formula
\begin{equation} \label{RhoT}
\rho =
1/G
-1/\text{f}(0) ,
\end{equation}
where $1/\text{f}(0)$ is the contact resistance ($\simeq 1$ at low
$T$). At zero temperature equation~(\ref{RhoT}) gives $\rho=\mathcal{R}
(\varepsilon_F)/\mathcal{T} (\varepsilon_F)$, where
$\mathcal{R}=1-\mathcal{T}$ and $\varepsilon_F$ is the Fermi energy.
Equations~(\ref{tra}-\ref{disprho}) were
derived~\cite{Landauer-70,Anderson-80,Mello-87,Anderson-81} by
averaging the formula $\rho = \mathcal{R} (\varepsilon_F)/\mathcal{T}
(\varepsilon_F)$. To extend that work to $T>0$~K means to average
eq.~(\ref{RhoT}) (see comment~\cite{comment1} for details).

We parametrize disorder by parameters $\mathcal{R}_I (k_F)$ and $N_I$,
where $\mathcal{R}_I(k)=\Omega^2/(k^2+\Omega^2)$ is the reflection
probability for a single $\delta$-barrier and $N_I$ is the 1D impurity
density. (Note that the final results do not depend on the choice of
$\mathcal{R}_I(k)$ as we assume very small $k_BT/ \varepsilon_F$. For
$\mathcal{R}_I (k_F)\ll 1$ we can also ignore fluctuations of
$\mathcal{R}_I (k_F)$ from impurity to impurity.) As the positions
$x_j$ are mutually independent, the distances $a=x_{j+1}-x_j$ between
the neighboring impurities in a given wire are selected at random from
the distribution $P(a)=N_I\exp[-N_Ia]$.

\begin{figure}[t]
\centerline{\includegraphics[clip,width=7.8cm]{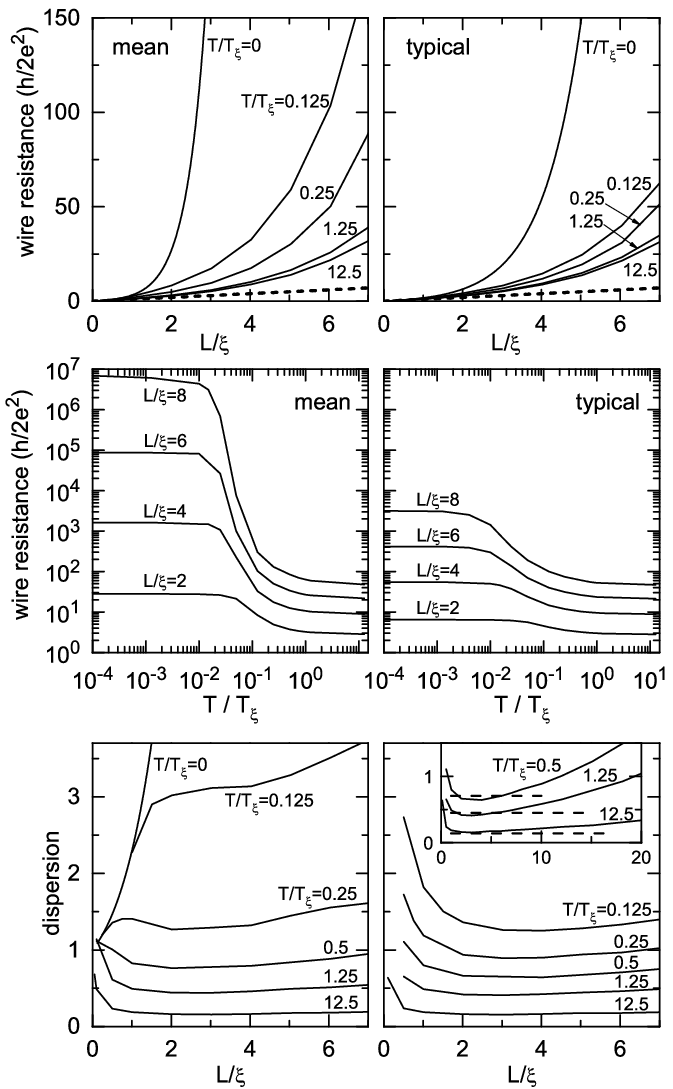}}
\caption{Top panel: Mean resistance $\left< \rho \right>$ and typical
resistance $\rho_t$ vs normalised length $L/ \xi$ for various
normalised temperatures $T/T_{\xi}$. The dashed line is the metallic
dependence $\rho = L/\xi$. Middle panel: $\left< \rho \right>$ and
$\rho_t$ vs $T/T_{\xi}$ for various $L/ \xi$. Bottom panel: Dispersion
of the resistance (left) and conductance (right) vs $L/\xi$ for various
$T/T_{\xi}$. Inset shows comparison with the estimate
$0.5\sqrt{T_{\xi}/T}$ (dashed line) derived in the text.}
\label{WireR}
\end{figure}

In Fig.~\ref{WireR} we show the averaged 1D resistance versus the wire
length and temperature. The results were obtained for disorder with
parameters $\mathcal{R}_I (k_F)=0.01$ and $N_I=10^7$ m$^{-1}$ and for
the electron gas parameters typical of the GaAs wire: $m=0.067$~$m_0$
and $\varepsilon_F = 35$~meV.

In fact, for $\mathcal{R}_I (k_F)\ll 1$ and $N_I^{-1}\gg2\pi/k_F$ (weak
low-density disorder) our results depend only on the parameters $L/\xi$
and $T/T_{\xi}$, independently on the choice of $\mathcal{R}_I (k_F)$,
$N_I$, $m$, and $\varepsilon_F$. Here $k_BT_{\xi} = 1 /
(g(\varepsilon_F) \xi)$, where $g(\varepsilon_F)=1/(\pi \hbar v_F)$ is
the density of energy levels. For weak low-density disorder the
localisation length is just the elastic mean free path, i.e., $\xi=
{(N_I R_I(k_F))}^{-1}$~\cite{Anderson-80}. For the above parameters
$\xi = 10$~$\mu$m and $T_{\xi} \simeq 1$~K.

In the top panel of Fig.~\ref{WireR} we reproduce at $T/T_{\xi}=0$ the
exponential growth~(\ref{tra}) and~(\ref{rt}). However, at
$T/T_{\xi}>0$ both $\left< \rho \right>$ and $\rho_t$ tend to grow with
$L/\xi$ much slower than predicted by eqs~(\ref{tra}) and~(\ref{rt}).
At $T/T_{\xi}>1$ they exhibit up to $L/\xi \leq 2$ the metallic
dependence $\left< \rho \right>=\rho_t = L/\xi$ with a nonlinear
correction which is $\leq 0.4 L/\xi$. For larger $L/\xi$ this nonlinear
growth is still far much slower than $\exp(L/\xi)$ and in general not
$\exp(\text{const} \ L)$ (see Fig.3).

The bottom panel of Fig.~\ref{WireR} shows the dispersions
$\Delta_{\rho}$ and $\Delta_{g}$, where $g = 1/\rho$. At $T=0$ K,
$\Delta_{\rho}$ follows the exponential rise~(\ref{disprho}) and
$\Delta_{g}$ diverges. As $T$ increases, $\Delta_{\rho}$ and
$\Delta_{g}$ decrease below unity and variation with $L$ tends to
disappear, unlike other transport regimes in which $\Delta_{\rho}$ and
$\Delta_{g}$ always vary with $L$ in some typical way \cite{Datta-95}.

\begin{figure}[t]
\centerline{\includegraphics[clip,width=8.1cm]{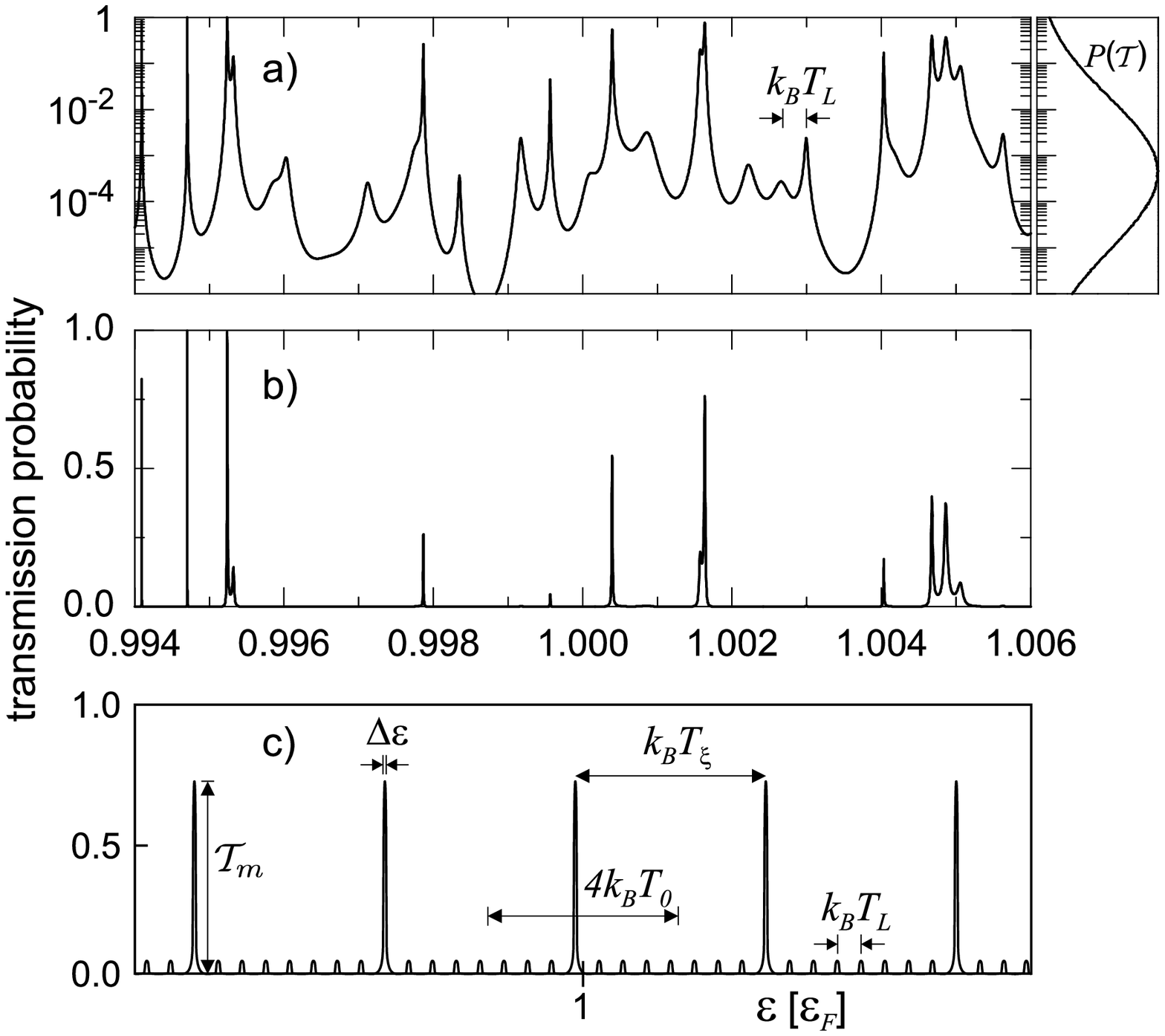}}
\caption{a) Transmission probability $\mathcal{T}(\varepsilon)$
versus energy $\varepsilon$ for one random configuration of
disorder and distribution $P (\mathcal{T})$ averaged over many
configurations. Energy axis is in units $\varepsilon_F$, where
$\varepsilon_F = 35$~meV. The length parameters are $L/ \xi=8$ and
$L=80$~$\mu$m, other parameters are the same as quoted for
figure~\ref{WireR}. b) The $\mathcal{T}(\varepsilon)$ dependence
from Fig.~(a) in linear scale. c) Schematic generalisation of
Figs. (a-b) (see the text).} \label{TE}
\end{figure}

To provide insight we now approximate eq.~(\ref{gT}) by
\begin{equation} \label{TEAprxE}
G \simeq \frac{1}{4k_{\text{B}}T} \int
_{\varepsilon_F - 2k_{\text{B}}T} ^{\varepsilon_F +
2k_{\text{B}}T} d\varepsilon \, \mathcal{T}(\varepsilon).
\end{equation}
A typical $\mathcal{T}(\varepsilon)$ dependence is shown in
Fig.~\ref{TE}. It consists of narrow peaks separated on average by
energy $k_BT_{L} = 1/(g(\varepsilon_F)L)$. Some of these peaks have
maximum close to unity due to the resonant tunneling across disorder,
the rest are the much lower peaks with a negligible area compared to
the area below the resonant peaks. The mean distance between the
resonant peaks is $k_BT_{\xi}$ and the mean number of such peaks in the
energy window $4k_{\text{B}}T$ is $4k_{\text{B}}T /
k_{\text{B}}T_{\xi}$~\cite{Azbel-84}. Thus, for $T \gtrsim T_{\xi}$ we
can estimate the mean and variance of the integral~(\ref{TEAprxE}) as
\begin{equation} \label{TEAprxE2}
\left< G \right> \simeq \frac{\mathcal{T}_m \, \Delta \varepsilon}{4k_{\text{B}} T}
\frac{4k_{\text{B}}T}{k_{\text{B}}T_{\xi}} , \ \
\sqrt{\text{var} (G)} \simeq \frac{\mathcal{T}_m \, \Delta \varepsilon}{4k_{\text{B}} T}
\sqrt{\frac{4k_{\text{B}}T}{k_{\text{B}}T_{\xi}}} ,
\end{equation}
where $\mathcal{T}_m$ and $\Delta \varepsilon$ are the average height
and width of the resonant peak. Equations~(\ref{TEAprxE2}) give the
$L$-independent dispersion $\Delta_{G} \simeq 0.5\sqrt{T_{\xi}/T}$, in
Fig.~\ref{WireR} compared with the simulated $\Delta_{g}$. A small
difference between $g$ and $G$ is not essential here, the estimate and
simulation differ because the $\text{var} (G)$ in eq.~(\ref{TEAprxE2})
ignores deviations from $\mathcal{T}_m \, \Delta \varepsilon$.

At $T \gtrsim T_{\xi}$ thermal averaging causes $\left< 1/G \right>
\simeq 1/\left< G \right>$. Then $\left< \rho \right> = \left< 1/G
\right> -1 \simeq 1/\left< G \right>$ for not too small $L/\xi$. To
illustrate how the coherent wire becomes more ``metallic'' than
insulating, we use in eq.~(\ref{TEAprxE2}) the ``ballistic'' maxima
$\mathcal{T}_m=1$ and $\Delta \varepsilon=k_BT_{L}$. We obtain $\left<
G \right> \simeq T_L/T_{\xi} = \xi/L$ and $\left< \rho \right> \simeq
L/\xi$. This ``metal'' exists if the window $4k_{\text{B}}T$ involves
at least one resonant peak. This happens for $4T/T_{\xi} \geq 1$, i.e.,
the crossover temperature is $T_0 = 0.25 T_{\xi}$.

For $T>T_0$ we see in the middle panel of Fig.~\ref{WireR} a tendency
to the $T$-independence, in accord with the $T$-independent $\left< G
\right>$ of eq.~(\ref{TEAprxE2}). Generally, $\left< \rho \right> =
1/\left< G \right> - 1$ for $T>>T_0$, but $\left< G \right>$ is
$T$-independent for any $T$:
\begin{equation} \label{gTaver}
\left< G \right> = \int _0 ^{\infty} d\varepsilon \ \left[-
  \frac{d\text{f}(\varepsilon)}{d\varepsilon} \right]
  \, \left< \mathcal{T}(\varepsilon) \right>  \simeq \int_{0}^{1}
d\mathcal{T} \mathcal{T} P (\mathcal{T}),
\end{equation}
where $P (\mathcal{T})$ is the distribution of transmission
$\mathcal{T}$ (Fig.~\ref{TE}).

\begin{figure}[t]
\centerline{\includegraphics[clip,width=8.6cm]{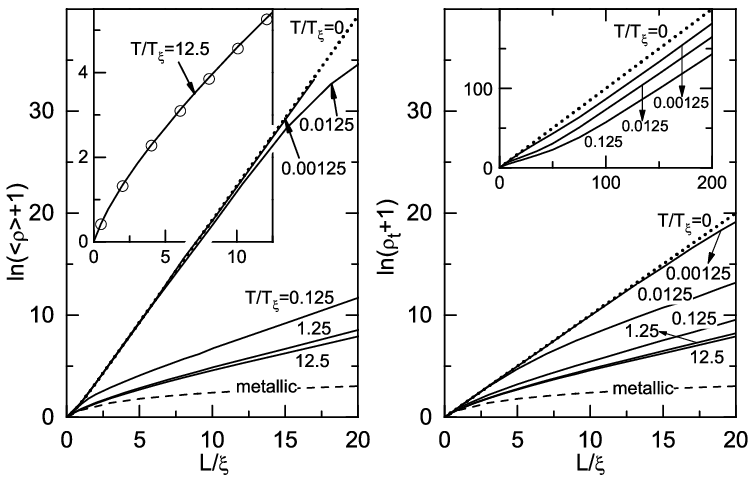}}
\caption{Mean resistance and typical resistance versus $L/\xi$ for
various $T/T_{\xi}$. The dashed line is the metallic resistance
$\rho=L/\xi$. Inset to the left panel: The $T = 12.5 T_{\xi}$ curve is
shown in more detail to stress the sublinear concave shape. The open
circles show the limit $\left< \rho \right> = {\left< G
\right>}^{-1}-1$, where $\left< G \right>$ is given by
eq.~(\ref{gTaver}) with $P (\mathcal{T})$ generated numerically. Inset
to the right panel: The typical resistance simulated up to $L/\xi=200$,
but only for $T / T_{\xi} \le 0.125$ owing to huge computational time.}
\label{lnRho}
\end{figure}

Figure~\ref{lnRho} shows $\left< \rho \right>$ and $\rho_t$ from
Fig~\ref{WireR} in semilog scale. For $T>T_{\xi}$ both $\left< \rho
\right>$ and $\rho_t$ rise with $L/\xi$ nonlinearly, but far much
slower than $\exp(L/\xi)$. Is this rise of the form $\exp(\text{const}
\ L)$? It is not at least for $L/\xi \lesssim 12$. Inset to the left
panel clearly shows that the rise is sublinear in semilog scale. Since
$\left< \rho \right> \rightarrow \rho_t$, we continue in terms of
$\rho_t$. It is defined by $\ln(1+\rho_t) \equiv \left< \ln(1+\rho)
\right>$. Thus, a physically allowed exponential form of $\rho_{t}(L)$
has to coincide with equation $\left< \ln(1+\rho) \right> =
\text{const} \times L$, which however cannot fit the sublinear curve in
the inset. The resistance thus rises with $L$ polynomially rather than
exponentially. This is neither the strong localisation (the SL is
manifested by eqs.~(\ref{tra} - \ref{disprho})) nor the weak
localisation (the WL occurs at $L/\xi \ll 1$), but somewhere between.
We therefore speak about the medium localisation, which means a
nonexponential decay of resonant transmission with $L/\xi$.

What happens for extremely large $L/\xi$? Inset to the right panel
shows the typical resistance for $T \ll T_{\xi}$. Here the rise
$\propto \exp(L/\xi)$ reappears at $L/\xi \gtrsim 50$. For $T>T_{\xi}$
neither $\rho_t$ nor $\left< \rho \right>$ is numerically feasible, but
for $L/\xi \gtrsim 40$ we find $\rho_t \simeq \left< \rho \right>
\propto (L/\xi)^{3/2} \exp(L/4\xi)$ by using other approach not
reported here. In the ``infinite'' wire also the resonant transmission
is finally damped exponentially.

Note however, that the limit of extremely long coherent 1D segments is
experimentally irrelevant. In general, the resistance $\gtrsim 10^9$
$\Omega$ is hardly measurable owing to intrinsic physical limitations
(e.g. the rf noise~\cite{Gershenson-97}). On the insulating side of the
crossover, one can thus measure $\rho_t \simeq e^{L/\xi}$ in principle
for $L/\xi \lesssim 12$. On the ``metallic'' side, both $\left< \rho
\right>$ and $\rho_t$ reach $10^9$ $\Omega$ for $L/\xi \simeq 30$ (not
shown in Fig.~\ref{lnRho}), so they could be measurable for $L/\xi
\lesssim 30$.

\begin{figure}[t]
\centerline{\includegraphics[clip,width=8.1cm]{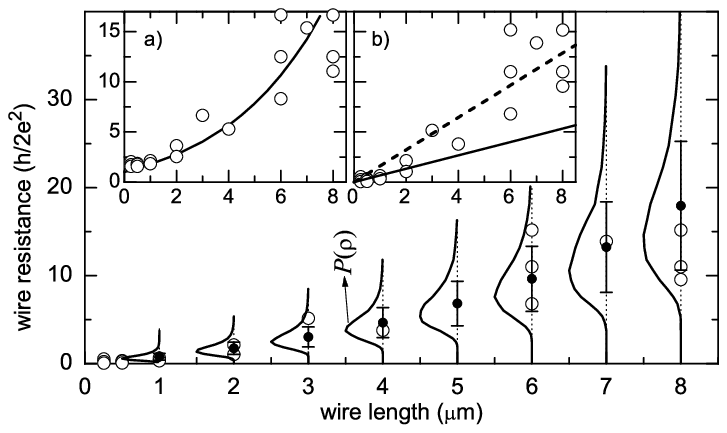}}
\caption{\textbf{a)} Open circles are the two-terminal resistance data
of Ref.~\onlinecite{Kaufman-99}, measured at $T=4.2$~K in a seria of
GaAs quantum wires with a single occupied channel. In these wires $\xi
\simeq 1.5$~$\mu$m and $\varepsilon_F \simeq 3.5$~meV, as the quantized
ballistic conductance is observed for $L \lesssim 1.5$~$\mu$m and the
lowest quantized step is centered roughly at the mentioned
$\varepsilon_F$ value~\cite{Kaufman-99}. For these parameters
$T_{\xi}\simeq 2$ K. The full line shows the simulated mean
two-terminal resistance, $\left< 1/G \right>$. In our simulation the
value $\xi \simeq 1.5$~$\mu$m can be realised through many different
combinations of parameters $N_I$ and $R_I(k_F)$, but the results remain
the same provided the disorder is weak. So we do not need to consider
sample-dependent details. \textbf{b)} Open circles are the experimental
data from inset (a) reduced by the contact resistance (the two-terminal
resistance at $L\rightarrow0$). To prove the superlinear rise with $L$
clearly, the experimental data are compared with the linear fit $\rho =
L/\xi$. The best fit is obtained for $\xi=0.6$~$\mu$m (dashed line).
However, it overestimates all experimental data for $L \le 2$~$\mu$m
and the value $\xi=0.6$~$\mu$m is in contradiction with the ballistic
conductance observed for $L \lesssim 1.5$~$\mu$m \cite{Kaufman-99}. For
$\xi=1.5$~$\mu$m (full line) the fit is indeed good for the ballistic
wire lengths, but for $L>2$~$\mu$m the experimental data grow
superlinearly. \textbf{Main panel:} The calculated mean resistance
$\left< \rho \right>$ (full circles), standard deviation from $\left<
\rho \right>$ (vertical bars), and resistance distribution $P (\rho)$
are shown. Open circles are the same as those in inset (b).}
\label{Kaufman}
\end{figure}

In Ref.~\onlinecite{Kaufman-99} the resistance of a single 1D channel
was measured for various $L>\xi$ in a seria of GaAs quantum wires
prepared by a method allowing to reproduce macroscopic parameters from
wire to wire. In Fig.~\ref{Kaufman} the experiment~\cite{Kaufman-99} is
compared with our simulation. In inset~(a) we compare the two-terminal
resistances. We see that the measured data reasonably reflect the
superlinear rise of the theoretical curve. Inset~(b) proves that the
superlinear rise of the measured data is systematically above $\rho =
L/\xi$ and cannot be ascribed to the data dispersion. The main panel
shows that the simulated mean resistance, standard deviation and
resistance distribution fit~\cite{comment1} the experiment. For a
perfect comparison experiments with very large ensembles of wires are
needed.

If decoherence would be present in the experiment \cite{Kaufman-99}, it
would cause the dependence $\rho = L/\xi$, which is not observed. This
means that in the measured 1D wires $L_{\phi} \gtrsim 8$~$\mu$m, which
is order of magnitude more than in a related 2D electron gas at the
same temperature~\cite{Datta-95}. Why is $L_{\phi}$ so large? At low
temperatures decoherence is due to the electron-electron interaction.
In the (ballistic) 2D electron gas the interaction of two electrons
fulfills the conservation laws $\varepsilon(k_1) + \varepsilon(k_2) =
\varepsilon(k'_1)+\varepsilon(k'_2)$ and $k_1+k_2=k'_1+k'_2$, which
allow the energy exchange and decoherence. In the 1D system such
conservation laws prohibit any energy exchange, which might be the
reason for large $L_{\phi}$. One might think~\cite{Imry-02} that
$L_{\phi} \lesssim \xi$ since the diffusion is controlled by
exponential localization. In our case $L_{\phi}$ can exceed $\xi$ many
times as the resistance of segment $L_{\phi}$ grows with $L_{\phi}/\xi$
weakly superlinearly, not exponentially. A full calculation of
$L_{\phi}$ in the 1D wire is beyond the Fermi-liquid theory and has not
yet appeared.

In summary, due to the thermal smearing and resonant tunnelling, the
disordered 1D insulator shows at full coherence the crossover to the 1D
``metal''. The resistance of the 1D ``metal'' grows with $L/\xi$ first
nearly linearly and then polynomially due to the medium localisation.
This is in contrast to the expectation that the resistance of the
coherent 1D system grows with $L/\xi$ exponentially if $L/\xi>1$. The
1D ``metal'' shows the resistance dispersion which is almost
$L/\xi$-independent and smaller than unity, again in contrast with the
expected (exponential) growth. Such 1D ``metal'' should be observable
in any coherent 1D system of length $L/\xi \lesssim 30$, longer
coherent segments are experimentally irrelevant. The crossover
temperature is $T_0=0.25T_{\xi}$, in the GaAs wires $T_0 \simeq
0.1-1$~K.

\begin{acknowledgments}
M. M was supported by the APVT Agency grant APVT-20-021602. P. V. was
supported by the EC fellowship HPMFCT-2000-00702 and VEGA grant
2/3118/23.
\end{acknowledgments}


\end{document}